\input{aipcheck}

\documentclass[
    ,final            
  ]
  {aipproc}

\layoutstyle{6x9}
\usepackage[T1]{fontenc}
\usepackage[utf8]{inputenc}
\usepackage{float}
\usepackage{amsmath}
\usepackage{graphicx}
\usepackage{amssymb}

\begin{document}

\title{Non-extensive radiobiology}


\author{O. Sotolongo-Grau}{
  address={UNED, Departamento de Física Matemática y de Fluidos}
}

\author{D. Rodriguez-Perez}{
  address={UNED, Departamento de Física Matemática y de Fluidos}
}

\author{J. C. Antoranz}{
  address={UNED, Departamento de Física Matemática y de Fluidos}
  ,altaddress={UH, Cátedra de Sistemas Complejos Henri Poincaré}
}

\author{O. Sotolongo-Costa}{
  address={UH, Cátedra de Sistemas Complejos Henri Poincaré}
}

\begin{abstract}
The expression of survival factors for radiation damaged cells 
is based on probabilistic assumptions and experimentally fitted for each
tumor, radiation and conditions. Here we show how the simplest of these 
radiobiological models can be derived from the maximum entropy principle
of the classical Boltzmann-Gibbs expression. We extend this derivation 
using the Tsallis entropy and a cutoff hypothesis, motivated by clinical 
observations. A generalization of the exponential, the 
logarithm and the product to a non-extensive framework, provides a 
simple formula for the survival fraction corresponding to the application
of several radiation doses on a living tissue. The obtained expression shows 
a remarkable agreement with the experimental data found in the literature, 
also providing a new interpretation of some of the parameters introduced anew. 
It is also shown how the presented formalism may has direct application in 
radiotherapy treatment optimization through the definition of the potential 
effect difference, simply calculated between the tumour and the surrounding tissue. 
\end{abstract}

\maketitle


\section{Introduction}

One of the main concerns of a radiation oncologist is to find a
treatment which, maximizing the damage over the tumor, minimizes it over the surrounding healthy tissue. In order to reach a suitable
treatment the radiobiologists have developed some empirical models
describing the interaction between radiation and living tissues (see
\cite{Tubiana} for a review of radiobiology models) capable of finding
the survival fraction, $F_{s}$, of cells under a radiation dose, $D$.
These models applicability limits are not clear so multiple corrections
have been developed in order to fit the experimental data \cite{Steel_ch8}.

The concept of tissue effect, $E$, raised from some of
these models \cite{Steel_ch10} is used to compare different
treatments each other. Usually expressed as $E=-\ln\left(F_{s}\right)$
is a dimensionless magnitude that gathers several models
of interaction between cells and ionizing radiation.

The simplest radiobiology model is the linear one. Here the tissue
effect is considered linear to the radiation dose, $E=\alpha D$,
and the survival fraction, $F_{s}=\exp\left(-\alpha D\right)$, is
viewed as the cumulative survival probability of a cell under any
dose below $D$. This probability fullfils the additive property meaning
that the effects of radiation are cumulative following an additive
model and the survival fraction for two doses could be found 
as $F_{s}\left[D_{1}+D_{2}\right]=F_{s}\left[D_{1}\right]\centerdot F_{s}\left[D_{2}\right]$.

However, this model only fits the experimental data for some tissues,
under low radiation doses \cite{Tubiana}, so the tissue effect must be corrected
to $E=\alpha D+\beta D^{2}$, called the linear quadratic (LQ) model. 
But then the survival fraction loses the additive
property, $F_{s}\left[D_{1}+D_{2}\right]<F_{s}\left[D_{1}\right]\centerdot F_{s}\left[D_{2}\right]$
, and the tissue effect becomes a supperadditive quantity, $E\left[D_{1}+D_{2}\right]>E\left[D_{1}\right]+E\left[D_{2}\right]$.

As a result of the nonlinear nature of $E$ in this case, the superposition
principle is not fulfilled. However any model of interaction between
radiation and living tissues must allow to divide a continuous radiation
in finite intervals and the resultant tissue effect must be the same.

Indeed, it is easy to show that, under the LQ viewpoint, if the tissue effect were additive
for different radiation sessions, then the additivity of the dose
would not hold. Conversely, assuming that the dose is additive then
the tissue effect is not equivalent to the sum of the effects
for different doses. This result suggests that the radiobiological
problem must be approached from a non extensive formulation \cite{tsallis-1999-29}.

In this work we use at the first stage the Boltzmann-Gibbs (BG) entropy
in order to find the expression of the tissue effect as a function
of the absorbed dose. Later, along with the Tsallis entropy \cite{Tsallis1991}
definition, it is assumed that a critical value of the radiation dose
kills every single cell and a general expression for survival fraction
is found. This survival fraction expression fits the experimental data
even where previous empirical models fail. Using the $q$-deformed functions
\cite{TSALLIS,Plastino99} a new expression to find the survival fraction
of a whole treatment is found allowing to show hints to find the best
treatment.

\section{First step: The classical approach}

First we study the extensive problem applying the BG entropy (in units
of the Boltzmann constant),\begin{equation}
S=-\int_{0}^{\infty}\ln[p(E)]p(E)dE,\label{eq:shannon}\end{equation}
where in this case $E$ is, as before, the tissue effect and $p(E)$
is the cell killing probability density. 

According to the maximum entropy principle, if $p(E)$ satisfies the
normalization condition and a finite mean value of the tissue effect
does exist, then the problem of finding the $p(E)$ that extremizes
the BG entropy under the above conditions can be posed. It is
well known that among all continuous probability distributions for
a positive continuous variable with a fixed mean value, the exponential
distribution has the largest entropy \cite{COVER}. So,\begin{equation}
p(E)=\frac{1}{\left\langle E\right\rangle }e^{-\frac{E}{\left\langle E\right\rangle }},\label{eq:BGE}\end{equation}
and the survival probability of a single cell will be\begin{equation}
F_{s}=\int_{E}^{\infty}p\left(x\right)dx=e^{-\frac{E}{\left\langle E\right\rangle }}.\label{eq:psBG}\end{equation}

The survival probability here must fulfill the dose additivity property.
This can be achieved if, following the discussion in the previous
section, the tissue effect is proportional to the absorbed dose:\begin{equation}
E=\alpha_{0}D,\label{eq:edef}\end{equation}
where $\alpha_{0}$ is chosen as a constant that makes $E$ adimensional.

It must be noted that \eqref{eq:psBG} is the experimentally proved
and currently used expression for the survival fraction as a function
of tissue effect and justified in the literature only through empirical
arguments \cite{Tubiana}. We can take $\alpha=\alpha_{0}/\left\langle E\right\rangle =1/\left\langle D\right\rangle $
and the expression \eqref{eq:psBG} becomes expressed in the known
standard radiobiology form of the linear model. 

Even when the BG treatment of the problem does not cover the available
data, it shows that the tissue effect must be defined as proportional
to the absorbed dose of radiation. However the empiric expressions
already known show, as has been discussed in the introduction, that the survival probability of a cell does not
fulfill the additive property. Since this is usually associated to
non extensive problems the solution must be searched using a non extensive
definition of entropy. On the other hand, the Tsallis formulation of the entropy has
been proved its helpfulness when applied to problems of this nature.

\section{One step further: The generalized approach}

To apply the maximum entropy principle, in the Tsallis version, to the
problem of finding the survival fraction of a living tissue \cite{Steel_ch5}
that receives a radiation, we postulate the existence of some amount
of absorbed radiation $D_{0}<\infty$ (or its equivalent ``minimal annihilation
effect'', $E_{0}=\alpha_{0}D_{0}$) after which no cell survives.
The application of the maximum entropy principle performs like the
usual one but with a few modifications.

The Tsallis entropy becomes\begin{equation}
S_{q}=\frac{1}{q-1}\left(1-\int_{0}^{E_{0}}p^{q}(E)dE\right)\label{eq:Tsallis_rent},\end{equation}
the normalization condition is in this case $\int_{0}^{E_{0}}p(E)dE=1$
and the $q$-mean value becomes $\int_{0}^{E_{0}}p^{q}(E)EdE=\left\langle E\right\rangle _{q}<\infty$.
With this definition, all properties of the tissue and its characteristics
of the interaction with radiation become included in $\left\langle E\right\rangle _{q}$
and therefore in $E_{0}$. This is the only parameter (besides $q$)
entering in our description. It is clear that the determination of
$\left\langle E\right\rangle _{q}$ for the different tissues under
different conditions of radiation would give the necessary information
for the characterization of the survival factor.

To calculate the maximum of \eqref{eq:Tsallis_rent} under the above
conditions the well known method of Lagrange multipliers \cite{Plastino99}
is applied, obtaining\begin{equation}
E_{0}=\frac{2-q}{1-q}\left(\frac{\left\langle E\right\rangle _{q}}{2-q}\right)^{\frac{1}{2-q}},\label{eq:omega}\end{equation}
and

\begin{equation}
p(E)=\left(\frac{2-q}{\left\langle E\right\rangle _{q}}\right)^{\frac{1}{2-q}}\left(1-\frac{1-q}{2-q}\left(\frac{2-q}{\left\langle E\right\rangle _{q}}\right)^{\frac{1}{2-q}}E\right)^{\frac{1}{1-q}}.\label{eq:lmT}\end{equation}

Then the survival factor is\begin{equation}
F_{s}(E)=\int_{E}^{E_{0}}p(x)dx=\left(1-\frac{E}{E_{0}}\right)^{\frac{2-q}{1-q}}\label{eq:psD},\end{equation}
with $q<1$ for $E<E_{0}$ and zero otherwise. It is not hard to see
that when $q\rightarrow1$ then $E_{0}\rightarrow\infty$ and $\left\langle E\right\rangle _{q}\rightarrow\left\langle E\right\rangle $.

Equation \eqref{eq:psD} can be written \begin{equation}
F_{s}(D)=\begin{cases}
\left(1-\frac{D}{D_{0}}\right)^{\gamma} & \forall D<D_{0}\\
0 & \forall D\geqslant D_{0}\end{cases}\label{eq:finalF},\end{equation}
where we introduced $E=\alpha_{0}D$, $\gamma=\frac{2-q}{1-q}$ and
$D_{0}=E_{0}/\alpha_{0}$. Finally, the LQ model is easily recovered from 
\eqref{eq:finalF} in the limit $q\rightarrow1$ up to order two in a Taylor
series expansion \cite{sotolongograu2009}.


\subsection{Tsallis based Survival fraction properties}

The linear model for the tissue effect \cite{Tubiana} implies that
if the dose is additive the corresponding survival fraction is multiplicative. 
Though this property belongs only
to the linear model and not to more general descriptions like the
LQ model \cite{Tubiana} and others, we think it is worth to find
a link between the additivity property of the dose and the probabilistic
properties of the cell survival fraction.

Let us define the $\exp_{\gamma}(x)$ function\begin{equation}
\exp_{\gamma}(x)=\left[1+\frac{x}{\gamma}\right]^{\gamma}\label{eq:expgdef},\end{equation}
and the $\ln_{\gamma}(x)$ its inverse function \begin{equation}
\ln_{\gamma}(\exp_{\gamma}(x))=x.\label{eq:lgdef}\end{equation}

Then, let us introduce the $\gamma$-product of two numbers $x$ and
$y$ as \begin{equation}
x\otimes_{\gamma}y=\exp_{\gamma}\left[\ln_{\gamma}(x)+\ln_{\gamma}(y)\right]=\left[x^{\frac{1}{\gamma}}+y^{\frac{1}{\gamma}}-1\right]^{\gamma}.\label{eq:gproddef}\end{equation}

Note that definitions \eqref{eq:expgdef} and \eqref{eq:lgdef} are
not essentially different from the $q$-exponential and $q$-logarithm
presented in \cite{TSALLIS}. We are just introducing these definitions
to simplify the calculations.

Let us now define the ``generalized tissue effect'' as $E=-\frac{E_{0}}{\gamma}\ln_{\gamma}(F_{s})$.
We demand this effect to satisfy the additive property. Then the survival
fraction expressed as
$F_{s}=\exp_{\gamma}(-\gamma\frac{E}{E_{0}})=\exp_{\gamma}(-\gamma\frac{D}{D_{0}}),\label{eq:fschi}$
becomes $\gamma$-multiplicative. This implies that the statistical
independence of the survival fractions is only possible when $\gamma\rightarrow\infty$
($q\rightarrow1$). 

The survival fraction for the sum of the effects after $N$ doses
becomes

\begin{equation}
F_{s}(NE)=\left[1-\sum_{i=1}^{N}\frac{E_{i}}{E_{0}}\right]^{\gamma}=\exp_{\gamma}\left[-\sum_{i=1}^{N}\gamma\frac{E_{i}}{E_{0}}\right]=\left[\bigotimes_{i=1}^{N}\right]_{\gamma}F_{s}(E_{i})\label{eq:superprod},\end{equation}
where $\left[\bigotimes_{i=1}^{N}\right]_{\gamma}$ denotes the iterated
application of the $\gamma$-product.

\section{Experimental agreement}

Equation \eqref{eq:finalF} represents the survival fraction in terms
of the measurable quantities $D$ (radiation dose) and $D_{0}$ (minimal annihilation dose). In
order to compare our model with the experimental data we have selected
some survival curves from the literature where the survival fraction
$F_{s}$ is represented as a function of $D$ for different radiation
conditions. However, if $D$ is rescaled as $1-D/D_{0}$, as usual
in phase transition phenomena, all curves corresponding to the same
tissue collapse to the same straight line in a log-log plot.

The expression of $\ln\left[F_{s}\right]$ has been fitted for 23 experimental
data sets, corresponding to 5 different tissues, in terms of the rescaled
variable $\ln\left[1-D/D_{0}\right]$, minimizing the appropriate least 
squares functional using the \emph{steepest descent} method \cite{NR}. 
The slope of these lines are the values of $\gamma$ meaning that $D_{0}$
is the natural unit of $D$.

\begin{figure}
\begin{centering}
\includegraphics[width=0.85\textwidth]{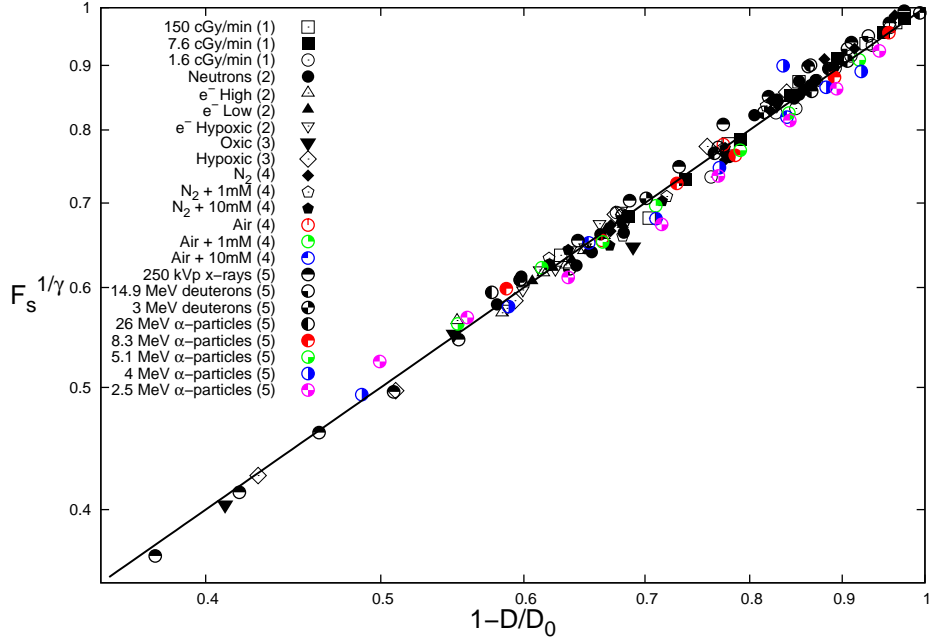}
\par\end{centering}

\caption{\label{fig:everybody}Normalized survival fractions, $\left(F_{s}\right)^{1/\gamma}$
, as a function of the rescaled radiation dose, $1-D/D_{0}$. The
straight line shown is $y=x$. Values for $\gamma$ and $D_{0}$
are detailed in table \ref{tab:fitting}.}

\end{figure}

Figure \ref{fig:everybody} shows, in a log-log plot, the comparison of our
model with all these data sets. In order to represent all data sets in
the same plot the survival fraction is shown normalized by $\gamma$ as 
$\left(F_{s}\right)^{1/\gamma}$.


\begin{table}[0.85\textwidth]
\begin{centering}
\begin{tabular}{lr}
\hline 
\tablehead{2}{p{0.9\textwidth}}{b}{(1) Human melanoma ($\gamma=14.0\pm0.9$) irradiated at different
dose rates. Data extracted from \cite{steel1987}.}\\
\hline 
$150\mbox{ cGy/min}$ & $D_{0}=27\pm2\mbox{ Gy}$\\
\hline 
$7.6\mbox{ cGy/min}$ & $D_{0}=38\pm4\mbox{ Gy}$\\
\hline 
$1.6\mbox{ cGy/min}$ & $D_{0}=46\pm5\mbox{ Gy}$\\
\hline 
\tablehead{2}{p{0.9\textwidth}}{b}{(2) Intestinal stem-cells ($\gamma=30.5\pm0.4$) irradiated with different
particles an conditions. Data extracted from \cite{alper1973}.}\\
\hline 
Neutrons & $D_{0}=36.0\pm0.7\mbox{ Gy}$\\
\hline 
Electrons (high dose rate) & $D_{0}=62.2\pm1.2\mbox{ Gy}$\\
\hline 
Electrons (low dose rate) & $D_{0}=68.8\pm1.6\mbox{ Gy}$\\
\hline 
Electrons (hypoxic conditions) & $D_{0}=162\pm3\mbox{ Gy}$\\
\hline 
\tablehead{2}{p{0.9\textwidth}}{b}{(3) Cultured mammalian cells ($\gamma=8.9\pm0.6$) exposed to x-rays
under oxic and hypoxic conditions. Data extracted from \cite{Hill1971}.}\\
\hline 
Oxic & $D_{0}=21.7\pm1.4\mbox{ Gy}$\\
\hline 
Hypoxic & $D_{0}=61\pm4\mbox{ Gy}$\\
\hline
\tablehead{2}{p{0.9\textwidth}}{b}{(4) Chinese hamster cells ($\gamma=14.2\pm0.6$) irradiated in the 
presence or absence of misonidazole. Data extracted from \cite{Adams1977}.}\\
\hline
Hypoxic & $D_{0}=85\pm7\mbox{ Gy}$\\
\hline 
Hypoxic, $1\textrm{mM}$ & $D_{0}=46\pm3\mbox{ Gy}$\\
\hline 
Hypoxic, $10\textrm{mM}$ & $D_{0}=33\pm2\mbox{ Gy}$\\
\hline 
Aerated & $D_{0}=29\pm3\mbox{ Gy}$\\
\hline 
Aerated, $1\textrm{mM}$ & $D_{0}=28\pm3\mbox{ Gy}$\\
\hline 
Aerated, $10\textrm{mM}$ & $D_{0}=32\pm4\mbox{ Gy}$\\
\hline 
\tablehead{2}{p{0.9\textwidth}}{b}{(5) Human kidney cells ($\gamma=8.8\pm0.3$) exposed in vitro to radiations 
of different energies. Data extracted from \cite{barendsen1968}.}\\
\hline
$250\textrm{kVp}$ x-rays & $D_{0}=22.4\pm0.8\mbox{ Gy}$\\
\hline 
$14.9\textrm{MeV}$ deuterons & $D_{0}=22\pm3\mbox{ Gy}$\\
\hline 
$3\textrm{MeV}$ deuterons & $D_{0}=17\pm2\mbox{ Gy}$\\
\hline 
$26\textrm{MeV}$ $\alpha$-particles & $D_{0}=15\pm2\mbox{ Gy}$\\
\hline 
$8.3\textrm{MeV}$ $\alpha$-particles & $D_{0}=9.1\pm1.2\mbox{ Gy}$\\
\hline 
$5.1\textrm{MeV}$ $\alpha$-particles & $D_{0}=7.5\pm0.8\mbox{ Gy}$\\
\hline 
$4\textrm{MeV}$ $\alpha$-particles & $D_{0}=6.9\pm0.6\mbox{ Gy}$\\
\hline 
$2.5\textrm{MeV}$ $\alpha$-particles & $D_{0}=9.2\pm0.7\mbox{ Gy}$\\
\hline
\end{tabular}
\par\end{centering}

\caption{\label{tab:fitting}Values for $\gamma$ and $D_{0}$ obtained from
the fitting with experimental data.}

\end{table}

All the information about the kind of radiation, radiation rate, etc.
is contained in the phenomenological term $D_{0}$, whereas tissues
are characterized by $\gamma$. This makes \eqref{eq:finalF} a very
general expression with universal characteristics since the phase
transition described by \eqref{eq:finalF} is homomorphic with the
phase transition of ferromagnets near the Curie point \cite{LANDAU}.
The exponent $\gamma$ in this case, as in ferromagnetic phase transitions,
determines the universality class. Then $\gamma$ in our case deals
only with the kind of tissue that interacts with radiation \cite{sotolongograu2009}.


\section{Is it possible to design the best protocol?}

The expression for the survival fraction after $N$ doses of radiation
must allow to compare the damage provoked on the tumor and surrounding
tissue cells if the magnitudes $\gamma$ and $D_{0}$ are known for
both, tumor and tissue. Let us assume we know the recommended radiation
dose $D_{+}$ per session for a treatment with $N$ sessions at several
values of $N$. We will define the potential effect as 

\begin{equation}
\chi=-\ln\left[F_{s}\right]\label{eq:chidef}, \end{equation}
and then the recommended potential effect per dose over the tumor
will be,\begin{equation}
\chi_{+}=-\gamma^{(t)}\ln\left[1-\frac{D_{+}}{D_{0}^{(t)}}\right]\label{eq:pot_total}, \end{equation}
where $\gamma^{(t)}$ and $D_{0}^{(t)}$ are the characteristic radiation
coefficients for the tumor. If we have already characterized
the healthy tissue around the tumor with $\gamma^{(h)}$ and $D_{0}^{(h)}$ then
we can find the survival fraction of cells after the treatment for
the tumor, \begin{equation}
\Theta^{(t)}=\exp_{\gamma^{(t)}}\left[N\ln_{\gamma^{(t)}}\left[\exp\left(-\chi_{+}\right)\right]\right]\label{eq:ft_tumor},\end{equation}
and the healthy tissue, \begin{equation}
\Theta^{(h)}=\exp_{\gamma^{(h)}}\left[\frac{\gamma^{(h)}D_{0}^{(t)}}{\gamma^{(t)}D_{0}^{(h)}}N\ln_{\gamma^{(t)}}\left[\exp\left(-\chi_{+}\right)\right]\right]\label{eq:ft_tejido}.\end{equation}

In order to find the best treatment all we need is to calculate the
difference of potentials for the treatment, \begin{equation}
\Delta\chi=\chi^{(t)}-\chi^{(h)}=\ln\left[\frac{\Theta^{(h)}}{\Theta^{(t)}}\right]\label{eq:dif_pot},\end{equation}
and guarantee that it will be positive. This could be achieved taking
advantage of the different responses to radiation of tumor and normal
tissues. Also, if this response is too close for an specific kind
of radiation or a given dose rate, those conditions could be changed
looking for a higher potential difference.

\section{Conclusions}

A new theoretical expression for the survival fraction of cells under
radiation has been found, using the Tsallis formulation of entropy.
The existence of a critical value for the absorbed radiation dose
under which no cells survive is introduced in the formulation in order
to get a proper expression. The new expression depends of two coefficients
that characterize the tissue behaviour under radiation ($\gamma$)
and the specifics conditions in which the radiation is applied ($D_{0}$).

The Tsallis mathematical formalism allows to redefine the multiplication
operation giving a way to find the survival fraction after several
radiation sessions. If the characteristic coefficients are known for
the tumor and the surrounding tissue then some hints can be given
to choose the less harmful, albeit most efficient, treatment to apply.


\begin{theacknowledgments}
The authors wish to thank Prof. Juan Antonio Santos, MD, for fruitful 
discussions. They also acknowledge the Spanish Ministerio de Industria 
for its support through the Proyecto CD-TEAM, CENIT. 
\end{theacknowledgments}



\bibliographystyle{aipproc}   

\bibliography{osotolongo}

\IfFileExists{\jobname.bbl}{}
 {\typeout{}
  \typeout{******************************************}
  \typeout{** Please run "bibtex \jobname" to optain}
  \typeout{** the bibliography and then re-run LaTeX}
  \typeout{** twice to fix the references!}
  \typeout{******************************************}
  \typeout{}
 }

\end{document}